\documentclass[aps,preprint,showpacs,preprintnumbers,amsmath,amssymb]{revtex4}
\begin{document}
\title{Coulomb scattering for scalar field in Scr\" odinger picture}

\author{Cosmin Crucean}
 \email{crucean@quasar.physics.uvt.ro}
\author{Radu Racoceanu}
\author{Adrian Pop}
 \email{apop302@quantum.physics.uvt.ro}
\affiliation{West University of Timi\c soara,  V. Parvan \\
 Avenue 4 RO-300223 Timi\c soara,  Romania}

\begin{abstract}
The scattering of a charged scalar field on Coulomb potential on de Sitter space-time is studied using the solution of the free Klein-Gordon equation. We find that the scattering amplitude is independent of the choice of the picture and in addition the total energy is conserved in the scattering process.
\end{abstract}

\pacs{04.62.+v}
\maketitle
\section{\label{sec:1}Introduction}

Recently a new time-evolution picture  was
defined in charts with spatially flat Robertson-Walker metrics,  under the name
of Scr\" odinger picture \cite{B1}.  Using the advantage offered by this
picture in \cite{B2} was found a new set of solutions for Klein-Gordon equation on de Sitter space-time which
behaves as polarized plane waves with a given energy . In the present paper we will exploit this, for calculate the Coulomb scattering for a charged scalar field in de Sitter expanding universe.

The paper is organized as follows.  In section \ref{sec:2},  we present a short review of
the Schr\" odinger picture introduced in \cite{B1} and we write the form of the
solutions for Klein-Gordon equation.  In section \ref{sec:3} we define the lowest
order contribution for scalar field in the potential $A^{\hat{\mu}}$ in
the new Schr\" odinger picture and then we calculate the scattering amplitude,
showing that the total energy is conserved. Our conclusions are summarized in section \ref{sec:4}. We use elsewhere natural units i. e.  $\hbar=c=1$.

\section{\label{sec:2}Plane waves with a given energy}

Let us take the local chart with cartesian coordinates of a flat
Robertson-Walker manifold,  in which the line element reads:
\begin{equation}\label{1}
ds^{2}=dt^{2}-\alpha(t)^{2}d\vec{x}\,^{2},
\end{equation}
where $\alpha$ is an arbitrary function. The scalar field $\phi$ of mass $m$ satisfy the free Klein-Gordon equation. If $\phi(x)$ is the scalar field in the natural picture then the scalar field of the Schr\"
odinger picture can be written as $\phi_{S}(x)=W(x)\phi(x)$ \cite{B2}, where $W(x)$ is the operator of time dependent dilatations \cite{B1},
\begin{equation}
W(x)=exp\left[-\ln(\alpha(t))(\vec{x}\cdot\vec{\partial})\right],
\end{equation}
which has the remarkable property:
\begin{equation}\label{w}
W(x)^{+}=\sqrt{-g(t)}W(x)^{-1}\,.
\end{equation}

Now taking in Eq. (\ref{1}) $\alpha(t)=e^{\omega t}$ one obtains the de Sitter
metric which is the case of interest here.
We start with the solutions of the free Klein-Gordon equations written on de Sitter space-time which was obtained in \cite{B2} using the Schr\" odinger picture. The fundamental solutions in the Schr\" odinger picture $f^S_{E,{\bf n}}$ of positive frequencies, with energy $E$ and momentum direction ${\bf n}$ resulted from \cite{B2} have the integral representation:
\begin{equation}\label{sol1}
f^S_{E,{\bf n}}(x)=\frac{1}{2}\sqrt{\frac{\omega}{2}}
 \frac{e^{-iEt}e^{-\pi k/2}}{(2\pi)^{3/2}}\int_{0}^{\infty} ds \sqrt{s}H^{(1)}_{ik}(s)
 e^{i{\bf p}\cdot{\bf x}-i\epsilon \ln s},
\end{equation}
where $s=p/\omega$ and $H^{(1)}_{\mu}(z)$ is a Hankel function of first kind .
Then the solution in the natural picture obtained in \cite{B2} read :
\begin{equation}\label{sol2}
f_{E,{\bf n}}({x})=\frac{1}{2}\sqrt{\frac{\omega}{2}}
 \frac{e^{-iEt}e^{-\pi k/2}}{(2\pi)^{3/2}}
\int_{0}^{\infty} ds\sqrt{s}H^{(1)}_{ik}(s)
  e^{i{\bf p}\cdot{\bf x}_t-i\epsilon \ln s},
\end{equation}
with ${\bf x}_t=e^{\omega t}{\bf x}$.
\par These solutions satisfy
the orthonormalization relations:
\begin{eqnarray}
i\int d^3x\, (-g)^{1/2}\, f_{E,\bf n}^*(x)
\stackrel{\leftrightarrow}{\partial_{t}} f_{E',{\bf n}'}(x)&=&\nonumber\\
-i\int d^3x\, (-g)^{1/2}\, f_{E,\bf n}(x)
\stackrel{\leftrightarrow}{\partial_{t}} f^*_{E',{\bf n}'}(x)
&=&\delta(E-E')\,\delta^2 ({\bf n}-{\bf
n}^{\,\prime})\nonumber\\
i\int d^3x\, (-g)^{1/2}\, f_{E,\bf n}(x)
\stackrel{\leftrightarrow}{\partial_{t}} f_{E',{\bf n}'}(x)&=&0
\end{eqnarray}
where the integration extends on an arbitrary hypersurface
$t=const$ and $(-g)^{1/2}=e^{3\omega t}$,
and the completeness condition
\begin{equation}
i\int_0^{\infty}dE\int_{S^2} d\Omega_n \left\{ [f_{E,{\bf n}}(t,{\bf
x})]^*\stackrel{\leftrightarrow}{\partial_{t}} f_{E,{\bf n}}(t,{\bf x}')
\right\} =e^{-3\omega t}\delta^3 ({\bf x}-{\bf x}^{\,\prime})\,.
\end{equation}

Since one study of the scattering in the energy basis was not done, we will focus in this paper on this problem. Our strategy is to obtain the scattering amplitude in the  Schr\" odinger picture and then to translate our calculations in the natural picture.

\section{\label{sec:3}Scattering amplitude and cross section}

The scattering theory in de Sitter spacetime can be constructed using the methods from flat space case. Then in analogy with Minkowski case \cite{B4} the scattering amplitude for a charged scalar field in the first order of the perturbation theory can be defined as follows:
\begin{equation}\label{ampl1}
A_{i\rightarrow f}=-e
\int\sqrt{-g(x)}\left[f^*_{f}(x)\stackrel{\leftrightarrow}{\partial_{\mu}}f_{i}(x)\right]A^{\mu}(x)d^{4}x,
\end{equation}
this being just the scattering amplitude in the natural picture.
It is then a simple calculations to obtain the scattering amplitude in the new Schr\" odinger picture using Eq. (\ref{w}), and we obtain:
\begin{equation}\label{ampl2}
A^{S}_{i\rightarrow f}=-e
\int \left[f^{S*}_{f}(x)\stackrel{\leftrightarrow}{\partial_{\mu}}f^{S}_{i}(x)\right]A^{\mu}_{S}(x)d^{4}x.
\end{equation}
If we write for the Coulomb potential on de Sitter space $A^{\hat{0}}_{S}(x)=\frac{Ze}{|\vec{x}|}e^{-\omega t}$,then in the new Schr\" odinger picture this will become:
\begin{equation}\label{12}
A^{\hat{0}}_{S}(x)=\frac{Ze}{|\vec{x}|}\,.
\end{equation}

Our aim here is to calculate the amplitude of Coulomb scattering for a charged scalar field using
the definition (\ref{ampl2}) in which we replace the solutions (\ref{sol1}) and the potential (\ref{12}).  Let us starting with the waves freely propagating in the $in$ and $out$ sectors, $f^{S}_{E_{i}, \bf{n}}(x)$ and $f^{S}_{E_{f}, \bf {n}}(x)$, assuming that the both of them are of positive frequency.  If we replace these quantities and perform the bilateral derivative, we observe that the spatial and temporal integrals have the same form as in Minkowski theory, and we obtain for our amplitude:
\begin{eqnarray}
A^{S}_{i\rightarrow f}=&&\frac{i\alpha Z \omega(E_{f}+E_{i})}{8\pi|\vec{p_{f}}-\vec{p_{i}}|^{2}}
\,\delta(E_{f}-E_{i})\left[\int_0^{\infty} ds_{f}
s^{1/2+iE_{f}/\omega}_{f}H^{(2)}_{ik}(s_{f})\right.\nonumber\\
&&\left.\times\int_{0}^{\infty}
ds_{i}s^{1/2-iE_{i}/\omega}_{i}
H^{(1)}_{ik}(s_{i})\right],
\end{eqnarray}
where $\alpha=e^{2}$.

The form of integrals that help us to obtain the final form of scattering amplitude are given in Appendix, here we give just the final result in terms of gamma Euler functions and delta Dirac function:
\begin{eqnarray}\label{ampl3}
A^{S}_{i\rightarrow f}= \frac{i\alpha Z\omega(E_{f}+E_{i})}{4\pi|\vec{p}_{f}-\vec{p}_{i}|^{2}}\,
\delta(E_{f}-E_{i})
\left[g_{k}(E_{f})g^{*}_{-k}(E_{i})\right].
\end{eqnarray}
For simplification in (\ref{ampl3}), we introduce the following notation:
\begin{eqnarray}\label{g}
 g_{k}(E)=&&
2^{iE/\omega}\frac{\Gamma(\frac{3}{4}+\frac{i
k}{2}+\frac{i E}{2 \omega})}{\Gamma(\frac{1}{4}+\frac{i
k}{2}-\frac{i E}{2
\omega})}-\frac{i2^{iE/\omega}}{\pi}\sin\left(\frac{\pi}{4}-\frac{ik
\pi}{2}+\frac{i E \pi}{2 \omega}\right)\nonumber\\
&&\times \Gamma\left(\frac{3}{4}+\frac{i k}{2}+\frac{i E}{2
\omega}\right)\Gamma\left(\frac{3}{4}-\frac{i k}{2}+\frac{i E}{2
\omega}\right),
\end{eqnarray}
and $g_{-k}(E)$ is obtained when $k \rightarrow -k$ in (\ref{g}).

Let us see now how the above results can be translated in the natural picture. The definition of the scattering amplitude in the natural picture is given in (\ref{ampl1}) and the form of Coulomb potential in the natural picture is $A^{\hat{0}}(x)=\frac{Ze}{|\vec{x}|} e^{-\omega t}$ . This time in our calculations we use the solutions of free Klein-Gordon equation written in natural picture (\ref{sol2}). After a little calculation one can observe that the result is the same as in (\ref{ampl3}). This can be checked passing to a new variable of integration $y=xe^{\omega t}$ when we solve the spatial integrals. It means that the result for the scattering amplitude is independent of the choice of picture in which one works.

Now let us make some comments about our scattering amplitude (\ref{ampl3}). We
obtain that the energy is conserved in the scattering process as
in Minkowski case. This result is expected because the form of
the external field (\ref{12}) allows us to consider that the
scattering process take place in a constant field. But it is known that
the energy of a system scattered on a constant field  is conserved \cite{B9}
(this does not mean that the momentum is also conserved), as we
obtained here.

 As we know the cross section can be defined in this case as \cite{B6}:
\begin{equation}\label{18}
d\sigma=\frac{dP}{dt}\frac{1}{j},
\end{equation}
where $\frac{dP}{dt}$ is the transition probability in unit of time and $j$ is the
incident flux.

The incident flux can be obtained using the definition of scalar current of particles . In the present case the incident flux calculated in the Schr\" odinger picture is:
\begin{eqnarray}\label{flux}
j=-i[f^*_{E_{i}\bf n}(x)\stackrel{\leftrightarrow}{\partial_{i}}f_{E_{i}\bf n}(x)]=\frac{ p_{i}\omega}{2 (2\pi)^{3}}g_{k}(E_{i})g^{*}_{-k}(E_{i}).
\end{eqnarray}
The evaluation of the probability in unit $\frac{dP_{l}}{dt}=\frac{d|A^{S}_{i\rightarrow
f}|^{2}}{dt}\frac{d^{3}p_{f}}{(2\pi)^{3}}$  of time is immediate and we obtain:
\begin{equation}\label{pl}
\frac{dP_{l}}{dt}=\frac{(\alpha Z)^{2}\omega^{2}(E_{f}+E_{i})^{2}}{32
\pi^{3}|\vec{p}_{f}-\vec{p}_{i}|^{4}}\delta(E_{f}-E_{i})
|g_{k}(E_{f})|^{2}|g_{-k}(E_{i})|^{2}.
\end{equation}
The differential cross section will be obtained after we replace (\ref{flux}) and (\ref{pl}) in (\ref{18}):
\begin{equation}\label{sec}
d\sigma=\frac{\omega(\alpha Z)^{2}\delta(E_{f}-E_{i})(E_{f}+E_{i})^{2}}{16p_{i}
\pi^{3}|\vec{p}_{f}-\vec{p}_{i}|^{4}}\frac{[|g_{k}(E_{f})|^{2}|g_{-k}(E_{i})|^{2}]}
{g_{k}(E_{i})g^{*}_{-k}(E_{i})}d^{3}p_{f}.
\end{equation}

As we know in Minkowski case the factor with $\delta(E_{f}-E_{i})$ was eliminated after
performing the integral with respect to the final momentum. In the present case an integration with respect final momentum is required for obtaining explicitly the differential cross section. However it is important to say that because we don't have a relation that connect the momentum and energy as in Minkowski case the distributional factor can not be eliminated when we perform the integration with respect to final
momentum in (\ref{sec}).

 We observe that our cross sections have a complicated dependence of
 energy. This dependence of energy was obtained after the integration with respect to $s=p/\omega$,
which means that this dependence translated in physical terms means that our cross section is dependent of the form of the incident wave which is unusual.

\section{\label{sec:4}Conclusion}

We examined in this paper the Coulomb scattering of a charged massive scalar field on de Sitter spacetime using
the plane wave solutions of free Klein-Gordon equation. The solutions that we use were obtained in  Schr\" odinger and natural pictures and behave as plane waves with a given energy. An important result is the fact that the scattering amplitude is independent of the picture in which we work.

 We found that the scattering amplitude and the cross sections depend
on the expansion factor as $\omega$. The result obtained here shows that the amplitude and the cross section depends on the form of the incident wave. Needless to say that this consequences is the result of the lost translational invariance with respect to time in de Sitter space-time. Also we found that the total energy is
conserved in the scattering process and, in addition, terms that
could break the energy conservation are absent since the
scattering was considered in a constant field of the form
(\ref{12}).

For further investigations it will be interesting to develop the entire scattering theory in Schr\" odinger picture. This will require an perturbation theory  and one reduction formalism for the scalar field, both developed in the new Schr\" odinger picture. For that one must use the form of the Klein-Gordon equation in the Schr\" odinger picture \cite{B2} and the fundamental solutions of positive/negative frequencies.

\section{Appendix}

For obtaining the scattering amplitude we need the formula \cite{B10}:
\begin{equation}
H^{(1,2)}_{\nu}(z)=J_{\nu}(z)\pm iY_{\nu}(z),
\end{equation}
which replaced in our amplitude led to integrals whose general form is \cite{B10}:
\begin{eqnarray}
\int_0^{\infty} dz
z^{\mu}J_{\nu}(z)=&&2^{\mu}\frac{\Gamma(\frac{\mu+\nu+1}{2})}{\Gamma(\frac{\nu-\mu+1}{2})},\nonumber\\
&&Re(\mu+\nu)>-1,\ Re(\mu)<\frac{1}{2}.
\end{eqnarray}
and
\begin{eqnarray}
\int_0^{\infty} dz
z^{\mu}Y_{\nu}(z)=&&\frac{2^{\mu}}{\pi}\Gamma\left(\frac{\mu+\nu+1}{2}\right)\Gamma\left(\frac{\mu-\nu+1}{2}\right)
\sin\frac{\pi}{2}(\mu-\nu),\nonumber\\
&&Re(\mu\pm\nu)>-1,\ Re(\mu)<\frac{1}{2}.
\end{eqnarray}
Now setting $z=p/\omega$ and $\nu=ik$ and $\mu=1/2\pm iE/\omega$ one can see that our
result (\ref{ampl3}) is correct. Also the above integrals help us to obtain the incident flux.

\end{document}